# Spin, orbital ordering and magnetic dynamics of LaVO$_3$: magnetization, heat capacity and neutron scattering studies

L D Tung[1*], A Ivanov[2], J Schefer[3], M R Lees[4], G Balakrishnan[4] and D M$^c$K Paul[4]

[1]*Department of Physics, University of Liverpool, L69 7ZE, United Kingdom.*

[2]*Institut Laue Langevin, BP156, 38042 Grenoble Cedex 9, France.*

[3]*Laboratory for Neutron Scattering, ETH Zürich & Paul Scherrer Institut, CH-5232 Villigen PSI, Switzerland.*

[4]*Department of Physics, University of Warwick, CV4 7AL, United Kingdom.*

We report the results of magnetization, heat capacity and neutron scattering studies of LaVO$_3$ single crystals. From the neutron diffraction studies, it was found that the compound is magnetically ordered with a C-type antiferromagnetic spin structure at about 136 K. In the vicinity of the ordering temperature, we also observed hysteresis in the neutron diffraction data measured on cooling and heating which indicates the first order nature of the phase transition. In the antiferromagnetically ordered phase, the inelastic neutron scattering studies reveal the presence of a temperature independent $c$ axis spin-wave gap of about 6 meV which is similar to that previously reported for the sister compound YVO$_3$.

## I. INTRODUCTION

The RVO$_3$ compounds (R = rare-earth or Y), often referred to as orthovanadates, are typical $t_{2g}$ electron systems that have attracted much attention recently due to their intriguing magnetic properties such as a temperature induced magnetization reversal,[1,2] magnetic memory effect,[3] staircase-like and glassy behavior in the magnetization[4] and a $c$ axis spin-wave gap in the magnetic dynamics of the vanadium spin modes.[5] In these compounds, the V$^{3+}$ ions have a $d^2$ electronic configuration with $S$ = 1 and these two electrons can occupy three possible orbital states $d_{xy}$, $d_{yz}$ and $d_{zx}$ giving rise to a strong interplay between the orbital and the spin degrees of freedom.

The crystal structure of the orthovanadates RVO$_3$ can be viewed as a stacking of the magnetically active VO$_2$ layers intercalated by RO layers. Long range orbital ordering (OO) and spin ordering (SO) in the vanadium sublattice have been observed at low temperature for different RVO$_3$ compounds. Early powder neutron diffraction studies revealed two possible magnetic structures.[6] Compounds with a larger rare earth radius (from La to Dy) have a C-type magnetic structure with antiferromagnetic (AF) spin order in the $ab$ plane (corresponding to a VO$_2$ layer) and ferromagnetic spin order along the $c$ axis. On the other hand, compounds with smaller rare earth ionic radii (from Ho to Lu) have a G-type magnetic structure with the vanadium spins ordered antiferromagnetically along the $a$, $b$ and $c$ axes. According to the Goodenough-Kanamori rules,[7] C- and G-type SO should correspond to G- and C-type OO, respectively.

Amongst the RVO$_3$ compounds, LaVO$_3$ has been studied extensively but there still exist a number of issues which have yet to be resolved. From the temperature dependence of the specific heat, it was suggested that on cooling, an AF SO takes place at a temperature T$_1$ (associated with a small anomaly in the heat capacity) and is followed by a structural transition at a temperature T$_2$ (corresponding to a peak in the heat capacity data).[8-10] The exact temperatures, T$_1$ and T$_2$, of these two transitions and their separation (approximately 2 K) vary from sample to sample. Structural studies by means of synchrotron X-ray measurements have confirmed that at T$_2$ there is an abrupt change in the lattice parameters associated with the first order crystallographic phase transition from orthorhombic *Pbnm* to monoclinic *P2$_1$/b11* symmetry with decreasing temperature.[9] Since the structural change from orthorhombic to monoclinic symmetry is usually accompanied by OO (see *e.g.* Ref. [11]), T$_2$ can also be identified as the OO temperature. On the other hand, at T$_1$, there is only a slight variation reported for the lattice parameters which does not result in any change in the crystallographic symmetry.

Because of the proximity in temperature of the phase transitions at T$_1$ and T$_2$, to date it has not been confirmed directly by neutron diffraction techniques whether T$_1$ or T$_2$ is truly related to the onset of the AF SO transition. Bordet *et al.*[12] studied LaVO$_3$ using both synchrotron X-ray and powder neutron diffraction techniques but they were unable to conclude whether or not the magnetic peaks appear at the same temperature as the monoclinic distortion or if the two transitions are separated by a few kelvin. It has been suggested that the magnetic structure of LaVO$_3$ in the narrow temperature region T$_2$ < T < T$_1$ might be of the G-type with the magnetic structure transforming into the C-type below T$_2$.[9] Thus, the evolution with temperature of the magnetic structure in LaVO$_3$ was proposed to be different to that seen in the other orthovanadate compounds (*e.g.* with R = Y, Dy, Er, Yb and Lu) where the magnetic structure is reported to

transform from the C-type to the G-type with decreasing temperature.[10]

There has also been speculation concerning the existence of a $c$ axis spin-wave gap in the magnetic dynamics of LaVO$_3$. This feature, which represents the energy difference between the optical and acoustic spin-waves at the magnetic zone boundary in the $c$ direction, was first observed by inelastic neutron scattering in YVO$_3$ in the C-type magnetic phase where it was measured to be about 5 meV.[5] To account for this gap, a scenario based on the existence of two different ferromagnetic exchange interactions, $J_c$, along the $c$ axis (i.e. dimerization) has been proposed. The nature of this dimerization was believed to be due to an orbital Peierls effect arising from the formation of an orbital dimer in which the thermal spin fluctuations play the role of a lattice degree of freedom.[5,13,14] On the other hand, based on the local density approximation and Hubbard energy calculations, it was shown that the spin-wave gap can also be reproduced by considering two different exchange interactions, $J_{ab}$, of the inequivalent VO$_2$ layers with different Jahn-Teller (JT) distortions in the $ab$ plane.[15] This latter model also predicted that such a gap should occur in the isostructural LaVO$_3$ compound but with a smaller magnitude of about 2 meV due to the reduced structural difference between the adjacent VO$_2$ layers.

In the following, we report the results of magnetization, heat capacity and neutron scattering studies on LaVO$_3$ single crystals. We have examined the nature of the magnetic order in this compound, paying particular attention to the possibility of a G-type magnetic ordering as suggested previously. Inelastic neutron scattering experiments have been carried out in order to explore the magnetic dynamics of the vanadium spin dispersive modes and to search for the predicted $c$ axis spin-wave gap in this compound.

## II. EXPERIMENTAL DETAILS

LaVO$_3$ single crystals have been grown by means of the floating zone technique according to the procedure described elsewhere.[2] The starting materials used for the crystal growth are La$_2$O$_3$ and V$_2$O$_5$ with purity of 99.9 %. Measurements of the temperature dependence of the magnetization were carried out in a Quantum Design SQUID magnetometer. Heat capacity measurements were performed in the temperature range from 2-300 K using a Quantum Design Physical Property Measurement System (PPMS). The magnetic structure of the compound was determined from single crystal neutron diffraction measurements at the TriCS instrument, Paul Scherrer Institute, Villigen, Switzerland, at a wavelength of 1.1807 (3) Å.[16] Special attention has been paid to the calibration of the temperature scales of the different instruments used for these studies. To this end, a reference LuVO$_3$ crystal which has a very sharp first order transition in the SO region[10] has been measured on the SQUID magnetometer, the PPMS calorimeter, and the TriCS diffractometer. The results (not shown) have indicated that the temperatures measured on all three instruments coincide to within experimental error which is less than 0.5 K. The temperature stability was estimated to be around 0.05 K or less during the neutron experiment using TriCS.

To study the dispersive magnetic excitations of the vanadium ions, we have also performed an inelastic neutron scattering experiment using the high flux thermal neutron triple-axis spectrometer IN8 at the Institut Laue Langevin, Grenoble, France. The instrument was operated with a double-focusing Si(111) monochromator and a Si analyzer to suppress the 2$^{nd}$ order harmonics at a fixed scattered wave vector of neutrons $k_f$ = 3.3 Å$^{-1}$.

We have selected and cut two crystals, C1 and C2, which were about 2 cm apart in the same as-grown rod. The mass of C1 is 13.66 mg and C2 113.9 mg. C1 was used in both the magnetization and the heat capacity measurements. As for C2, because of its relatively large size it was first used in the neutron diffraction measurements at TriCS, then, a smaller piece of 21.303 mg was cut from C2 for the magnetization and heat capacity measurements in order to compare the data with that obtained on C1. For the inelastic neutron scattering studies, which demand a relatively large volume of sample, we have grown a different crystal (C3) with a mass of about 2 g. Characterization by means of the X-ray Laue technique indicated that all the crystals were of a good quality. However, because of a very small difference between the $a$ and $b$ lattice parameters (~0.03 %), we were unable to rule out the possibility that the crystals were twinned using either X-ray Laue or neutron techniques. For the small crystal C2 used at TRICS, subsequent high resolution X-ray scans on the surface indicated no evidence of twinning.

## III. RESULTS

The crystallographic structure of LaVO$_3$ can be described within the orthorhombic space group *Pbnm*.[17] At 180 K, the lattice parameters derived from our single crystal neutron diffraction data for LaVO$_3$ are $a$ = 5.5178 (0.0151) Å, $b$ = 5.5149 (0.0105) Å, and $c$ = 7.7897 (0.0203) Å.

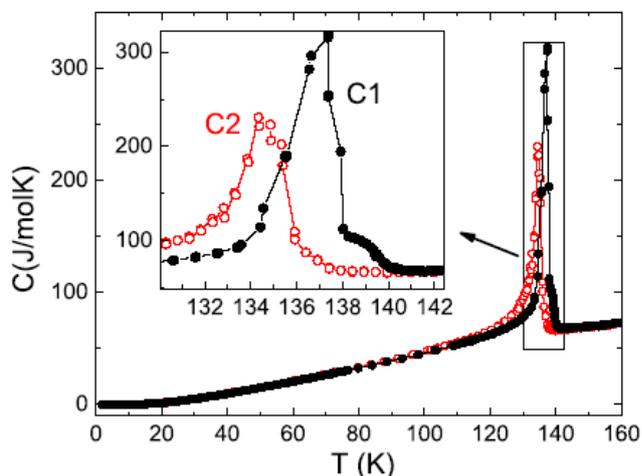

**Fig. 1** (color online): Heat capacity as a function of temperature for two different crystals C1 and C2 of LaVO$_3$. The inset shows the data near the peak.

In Fig. 1, we present the heat capacity data for the two crystals C1 and C2. Despite the fact that both crystals were cut from the same rod, certain differences are observed in the position and shape of the principal features in the temperature dependence of the specific heat. In C1, a clear peak at $T_2$ is followed closely by a shoulder on the high-temperature side at $T_1$. Such a behavior in the heat capacity of $LaVO_3$ was observed previously by other groups,[8-10] who attributed the peak and shoulder to the OO and SO temperatures, respectively. In the crystal C2, our data reveal only a slight change in the slope of the heat capacity data above the peak, and not a clear shoulder as in C1. The differences in the specific heat curves and the positions of the transition temperatures are probably due to differences in the stoichiometry of the two crystals.

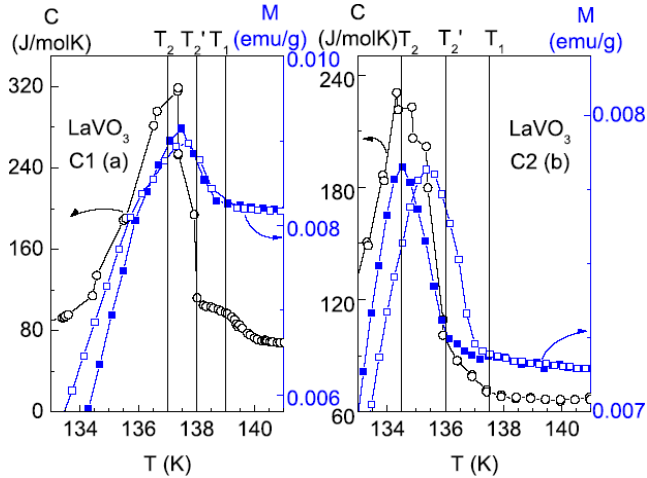

**Fig. 2** (color online): Heat capacity (left axis, circles) and magnetization (right axis, squares) as a function of temperature of two $LaVO_3$ single crystals C1 (a) and C2 (b). The open symbols correspond to data points collected on warming, the filled symbols to data points collected on cooling.

In Fig. 2, we plot the field-cooled cooling (FCC) and field-cooled warming (FCW) magnetization as a function of temperature for the two crystals in a narrow temperature window around the transitions. Here, the heat capacity data are also added to the Figure for comparison. In both crystals, the upturn in the FCC and FCW magnetization occurs at $T_1$ which corresponds to the shoulder in the heat capacity data for crystal C1 and the slight change in the slope in the heat capacity data for crystal C2. In Fig. 2, we have also marked the temperatures $T_2$ and $T_2$' where we observe the peak and the initial rise in the heat capacity data, respectively. There is a much larger temperature hysteresis in the FCC/FCW magnetization data for sample C2.

The temperatures $T_1$ and $T_2$ ($T_2$') observed in our two crystals are higher than those reported by Borukhovich *et al.* (135.7 and 133.4 K)[8] but lower than those reported by Ren *et al.*[9] and Miyasaka *et al.*[10] (143 and 141 K). While the difference in the observed values of $T_1$ and $T_2$ ($T_2$') could be attributed to different stoichiometries (*i.e.* small variations in either the La, V or oxygen content), the physics underlying these transitions should be the same for all the different $LaVO_3$ samples. Here, it is worth mentioning that the proximity of $T_1$ to $T_2$ ($T_2$') is not limited to $LaVO_3$ but can also be observed in Sr substituted $La_{1-x}Sr_xVO_3$ samples (x < 0.17) in which the transitions are shifted monotonously towards lower temperatures with increasing strontium content.[18]

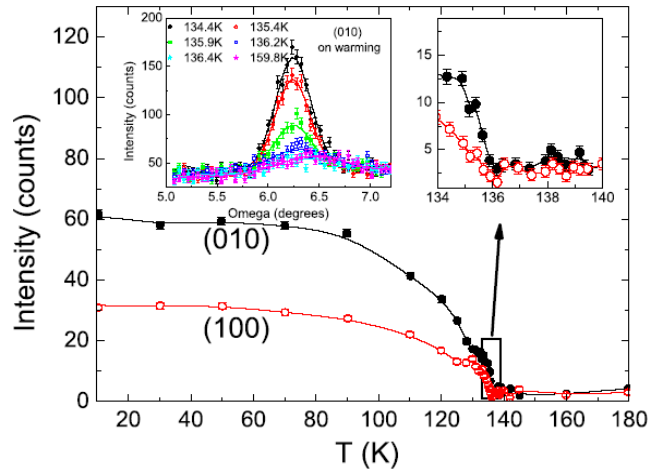

**Fig. 3** (color online): Integrated intensity of the (0 1 0) and (1 0 0) reflections measured on cooling a $LaVO_3$ single crystal (C2) as a function of temperature. The right hand inset shows a view around the transition at $T_2$'$^{cooling}$ ≈ 136 K. The left hand inset shows an omega scan of the (0 1 0) reflection measured on warming at some temperatures near the transition ($T_2$'$^{warming}$ ≈ 136.4 K).

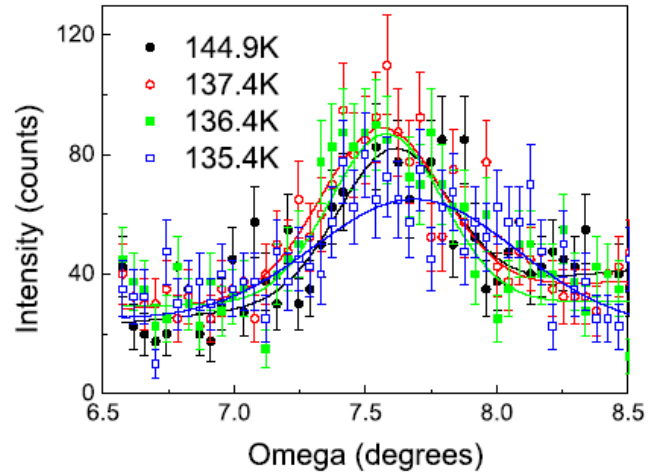

**Fig. 4** (color online): An omega scan of the (0 1 1) reflection of a $LaVO_3$ single crystal (C2) at some selected temperatures above $T_1$ = 134.5 K.

In order to correlate the changes in lattice and magnetic structure with the temperatures $T_1$ and $T_2$ ($T_2$') derived from the heat capacity and magnetization data, we have performed single crystal neutron diffraction measurements on the crystal C2. On cooling, we observe that immediately below 136 K ($T_2$' in Fig. 2b), a magnetic contribution to the reflections (h k l), where h is zero or even (odd), k odd (zero or even) and l zero or even, starts to develop, as illustrated in Fig. 3 for the (0 1 0) and (1 0 0) reflections. This set of magnetic Bragg reflections is characteristic of the C-type magnetic structure.[5] There is a visible hysteresis (about 0.4 K) in the value of $T_2$' as determined from the Bragg intensities between the cooling and warming measurements consistent with the magnetization data and indicating that the transition is first order (Fig. 3).

We paid particular attention to the possible occurrence of a G-type magnetic structure in the narrow temperature range between 134.5 K = $T_2$ < T < $T_1$ = 137.5 K, as proposed previously by Ren et al.[9] If a G-type magnetic structure does develop, additional magnetic intensity should have been seen on top of the (h k l) reflections with h zero or even (odd), k odd (zero or even), l odd.[5] However, our single crystal neutron diffraction data indicate that it is not the case. In Fig. 4, as an example, we plot the omega scan of a (0 1 1) reflection at some selected temperatures between 134.5 K = $T_2$ < T < $T_1$ = 137.5 K which can be compared with data measured at 144.9 K in the paramagnetic region. Within experimental error, we see no evidence for an enhancement in the intensity with respect to that seen at 144.9 K ruling out the development of a G-type SO below $T_1$ = 137.5 K.

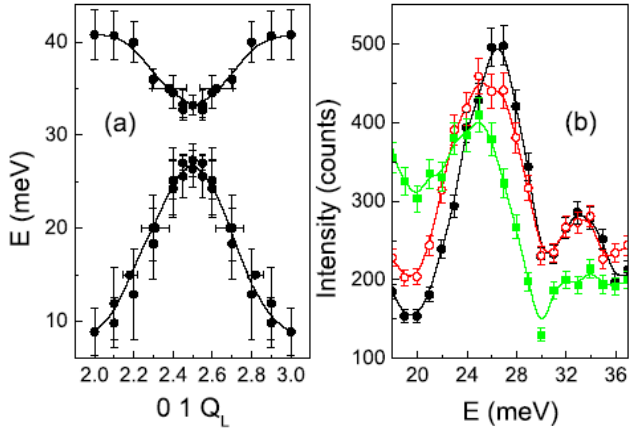

**Fig. 5** (color online): (a) Magnon dispersion relations at 2 K and (b) an energy scan along (0 1 2.5) at 2 K (●), 100 K (○) and 150 K (■) for a LaVO$_3$ single crystal (C3).

The results of our inelastic neutron scattering experiment on the magnon dispersion relations at 2 K along the (0 1 $Q_L$) direction are plotted in Fig. 5a. We observe a splitting of the dispersive V modes into optical and acoustic magnons with a gap of about 6 meV at the magnetic zone boundary ($Q_L$ = 2.5). This c axis spin-wave gap shows no significant variation between 2 and 100 K. At 150 K, in the paramagnetic region, the intensity is decreased and the signal is shifted to lower energies (see Fig. 5b). Our whole set of observations as a function of temperature and momentum transfer proves the magnetic origin of the signal depicted in Fig. 5. We would like to note that, in a similar inelastic neutron experiment, Ulrich and co-workers have observed no gap in the spin-wave dispersions of LaVO$_3$[19] which would be in agreement with the Goodenough-Kanamori rules.

## IV. DISCUSSION

Our single crystal neutron diffraction studies, show that the AF C-type SO in LaVO$_3$ occurs at the same temperature as the sharp upturn in the heat capacity at $T_2'$ = 136 K and that the compound does not order at any temperature with the G-type magnetic structure. Regarding the temperature at which the G-type OO occurs, we consider the three temperatures $T_1$ = 137.5 K, $T_2'$ = 136 K and $T_2$ = 134.5 K where "anomalies" are seen in our magnetization and heat capacity data (Fig. 2b). According to the previous studies by Ren et al.,[9] the first order structural phase transition that coincides with a G-type OO could give rise to either the peak in the heat capacity at $T_2$ = 134.5 K or to the upturn at $T_2'$ = 136 K. Within the resolution of our neutron diffraction measurements, it is difficult to detect directly any change in the structure of LaVO$_3$.[12] However, we observe hysteresis between cooling and warming measurements in the neutron and the magnetization data around $T_2'$ = 136 K. This hysteresis points to a first order phase transition that might be related to the structural change that coincides with the G-type OO. In other words, our observations allow us to suggest that the C-type AF SO and G-type OO occur at the same temperature $T_2'$ = 136 K.

Regarding the temperature $T_1$ = 137.5 K, where we observe the upturn in the FCC and FCW magnetization as well as a small 'anomaly' in the heat capacity data, the results of our single crystal neutron diffraction provide no evidence for either a C- or a G-type SO at this temperature. To date, these are the only two magnetic structures that have been observed in the different RVO$_3$ compounds (see e.g. Ref. [6]). Whilst $T_1$ seems to be accompanied by a small change in the lattice parameters,[9] we suggest that it may also be linked to the development of short range OO/SO which is difficult to detect in our neutron diffraction data.

The c axis spin-wave gap observed in the spin-wave spectrum of LaVO$_3$ as seen in Fig. 5 reveals unusual and unexpected features. Firstly, the (almost) temperature independence of the gap (Fig. 5b) seems to contradict the model based on orbital quantum fluctuations,[5,13,14] which claims that the origin of the c axis exchange bond dimerization is the orbital Peierls effect, in which thermal spin fluctuations play the role of a lattice degree of freedom. Secondly, the gap observed for LaVO$_3$ is about the same as that of YVO$_3$ even though the JT distortion in LaVO$_3$ is believed to be much weaker than in YVO$_3$.[20] Our results, therefore, do not appear to support the mechanism of two different exchange interactions $J_{ab}$ within the inequivalent VO$_2$ layers with different JT distortions, as proposed by Fang et al.[15] We note the fact that single crystal samples of LaVO$_3$ from different sources may not have exactly the same stoichiometry. This clearly influence the values of the ordering temperatures; the type and density of defects present in different LaVO$_3$ single crystals may also vary from sample to sample. The number of defects might be quite small, but in quasi one-dimensional orbital systems such as RVO$_3$, they may have a significant effect on the magnetic properties of the material. This occurs because the introduction of defects limits the growth of the correlation length along the orbital chains, dramatically influencing the long-range spin/orbital ordering and thus the magnetic response within each chain. To date, the effects of inhomogeneity and quenched disorder have not yet been treated theoretically for the RVO$_3$ compounds. Such effects have been considered in the closely related manganite RMnO$_3$ systems in order to explain the colossal magnetoresistance observed in this class of materials.[21] An inhomogeneity model based on defects can also be used to qualitatively account for some of the intriguing magnetic properties observed in the orthovanadate RVO$_3$ compounds such as the staircase-like and glassy behavior in the magnetization,[4] and the magnetization reversal.[2] Quenched disorder caused by defects in the orbitals of the quasi one-dimensional system could also play important role in the appearance or otherwise of the *c* axis spin-wave gap of the magnetic dynamics of vanadium modes in the $LaVO_3$ and $YVO_3$ compounds.

In summary, we have carried out magnetic, heat capacity and neutron scattering studies on $LaVO_3$ single crystals. We observe C-type SO but find no evidence for G-type SO in $LaVO_3$. Our studies of the magnetic dynamics of the vanadium dispersive modes in the SO regions have revealed the presence of temperature independent *c* axis spin-wave gap of about 6 meV. We suggest that the size of the gap may be modulated by the degree of quenched disorder in this inhomogeneous quasi one dimensional system.

This research project has been supported by the European Commission under the 6th Framework Programme through the Key Action: Strengthening the European Research Area, Research Infrastructures, contract RII3-CT-2003-505925. Part of the work is based on experiments performed at the Swiss spallation neutron source TRiCS/SINQ, Paul Scherrer Institute, Villigen, Switzerland and at Institute Laue Langevin, Grenoble, France. The financial support from EPSRC is gratefully acknowledged.

___________________________________________________

*email address: ltung@liv.ac.uk or leductung2000@gmail.com